\documentclass[12pt,onecolumn,journal]{IEEEtran}

\usepackage{cite}
\usepackage[cmex10]{amsmath}
\usepackage{subfigure}
\ifCLASSINFOpdf
  \usepackage[pdftex]{graphicx}
\else
  \usepackage[dvips]{graphicx}
  \graphicspath{{../figs/}}
  \DeclareGraphicsExtensions{.eps}
\fi

\begin{document}
\title{Cooperative Wideband Spectrum Sensing for the Centralized Cognitive Radio Network}
\author{Peng~Zhang,~\IEEEmembership{Student Member,~IEEE,}
        Robert~Qiu,~\IEEEmembership{Senior Member,~IEEE,}
\thanks{The authors are with the Wireless Networking Systems Lab in Department of Electrical and Computer Engineering, Center for Manufacturing Research, Tennessee Technological University, Cookeville, TN, 38505, USA. E-mail: pzhang21@students.tntech.edu, rqiu@tntech.edu}}

\maketitle
\begin{abstract}
Various primary user (PU) radios have been allocated into fixed frequency bands in the whole spectrum. A cognitive radio network (CRN) should be able to perform the wideband spectrum sensing (WSS) to detect temporarily unoccupied frequency bands. We summarize four occupancy features for the frequency bands. 1. The occupancy is sparse; 2. The frequency band allocation information is fixed and common; 3. There are three categories for the frequency band usages; 4. The occupied frequency bands are common in the CRN. For the first time, we consider all features as the prior knowledge in the compressed sensing based cooperative WSS (CWSS) algorithm design for a centralized CRN. We propose a modified orthogonal matching pursuit (Mod-OMP) algorithm and a modified simultaneous orthogonal matching pursuit (Mod-SOMP) algorithm for the CWSS. We compare the CWSS performance of Mod-OMP/Mod-SOMP with the original OMP/SOMP and show the performance improvements.
\end{abstract}

\section{Introduction}
\label{Introduction}
According to the US frequency allocation chart \cite{US_Freq_Allocation}, a variety of primary user (PU) radios have been allocated to the frequency bands covering the whole spectrum. However, existing PUs are not occupying all frequency bands all the time \cite{haykin2005cognitive}. For example, filed measurement of a $2.5$ GHz spectrum shows that the occupancy rate can be as low as $10\%$ \cite{bwrcspectrum}. Cognitive radio network (CRN) is a network of secondary user (SU) using the temporalily unoccupied frequency bands. A CRN should be able to perform the wideband spectrum sensing (WSS) to detect those unoccupied frequency bands.

One challenge for the WSS is that the possible unoccupied frequency bands are spread over a multi-GHz spectrum. If the WSS is performed following the Nyquist sampling rate, it will be infeasible for current analog-to-digital converters (ADC). Several features can be utilized to overcome the sampling rate limitation. Feature 1 is that the spectrum is sparse, which means that the spectrum occupancy rate is low. This has made the compressed sensing based algorithms possible for the WSS. Several compressed spectrum sensing (CSS) receivers \cite{AIC_2008,xampling_2009} have been proposed to perform the WSS under sub-Nyquist sampling rate. Feature 2 is that the frequency bands allocated to the PUs are fixed and can be considered as the prior knowledge to the SU. Reference \cite{liu2010compressive} has made use of such prior knowledge and proposed a basis pursuit (BP) \cite{chen2001atomic} based algorithm, the mixed $l_2/l_1$ norm denoising operator algorithm (MNDO), reaching a better reconstruction performance than the traditional reconstruction algorithms. However, our previous work in \cite{Zhang2010ModOMP} shows that when a third feature is considered, the reconstruction performance can be further improved. Feature 3 is that the usages of the frequency bands can be grouped into 3 categories \cite{haykin2005cognitive}:
\begin{itemize}
	\item Some frequency bands in the spectrum are largely unoccupied most of the time.
	\item Some other frequency bands are only partially occupied.
	\item The remaining frequency bands are heavily used.
\end{itemize}
In \cite{Zhang2010ModOMP}, we have demonstrated that feature 3 can be integrated into the classical greedy algorithm, orthogonal matching pursuit (OMP) \cite{tropp2007srr}, and we have proposed a modified OMP (Mod-OMP) reconstruction algorithm for the single user WSS. 

When multiple SUs can cooperate, the cooperative WSS (CWSS) is a common approach to achieve better WSS performance in a CRN. An additional feature needs to be pointed out for the CWSS: Feature 4, the occupied frequency bands are common among all SUs within the same CRN. The CRN can be centralized or decentralized. Here we consider the centralized CRN with a fusion center. All SUs can only cooperate with the fusion center to perform the WSS. Moreover, we consider the greedy algorithms due to its low computation complexity \cite{tropp2007srr}. To the best of our knowledge, the CWSS algorithms using greedy algorithms for the centralized CRN are very limited. In \cite{wang2009distributed}, a distributed compressed sensing (DCS) \cite{DCS2005} based architecture is proposed for the centralized CRN. It uses the simultaneous orthogonal matching pursuit (SOMP) reconstruction algorithm to perform the CWSS. However, it only takes advantage of feature 1 and feature 4. 

In this correspondence, we focus on the design of the CWSS algorithm using CSS for a centralized CRN. For the first time, all four features are integrated. Two common schemes for the CWSS are considered: decision fusion and data fusion \cite{zeng2010review}. For the decision fusion CWSS, we make use of the single user Mod-OMP and propose the Mod-OMP based decision fusion algorithm (Mod-OMP DeF). For the data fusion CWSS, we propose a novel modified SOMP (Mod-SOMP) algorithm as well as the Mod-SOMP based data fusion algorithm (Mod-SOMP DaF). We adapt the Hanssen-Kuiper skill score or R score \cite{chen2004statistical} metric to evaluate the accuracy of the decisions made upon the frequency band occupancy.
Simulations are conducted to compare the performances of the decision fusion algorithm with original OMP (OMP-DeF), the data fusion algorithm with original SOMP (SOMP-DaF), as well as the proposed Mod-OMP DaF and the Mod-SOMP DeF. 

This correspondence is organized as follows. In Section \ref{System_Model}, we review the system models of the CSS and the R score metric. In Section \ref{CWSS_Algorithms}, we introduce the four CWSS algorithms in detail. Simulation results for all algorithms are shown in Section \ref{Simulation_Results} and conclusions are made in Section \ref{Conclusions}.
\section{System Model}
\label{System_Model}
\begin{figure}[ht]
	\begin{minipage}[b]{.5 \linewidth}
	\centering
		\includegraphics[width=1.0\textwidth]{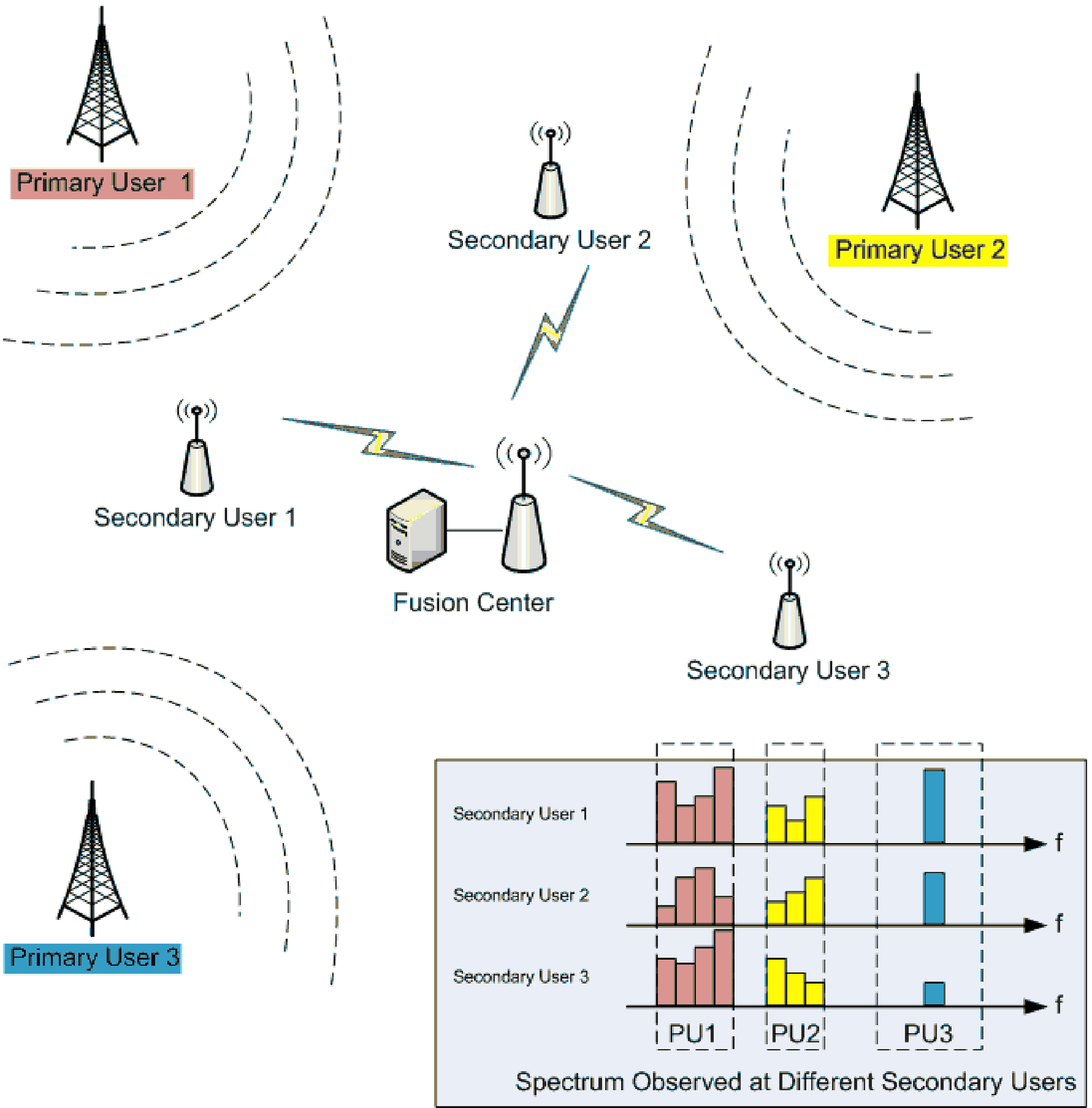}
	\caption{Illustration of a typical CRN coexisting with several PUs. The spectrum measured at all SUs have common non-zero entries.}
	\label{fig:Dist_Spec_Sensing}
	\end{minipage}
\hspace{.5cm}
	\begin{minipage}[b]{.5 \linewidth}
	\centering
		\includegraphics[width=1.0\textwidth]{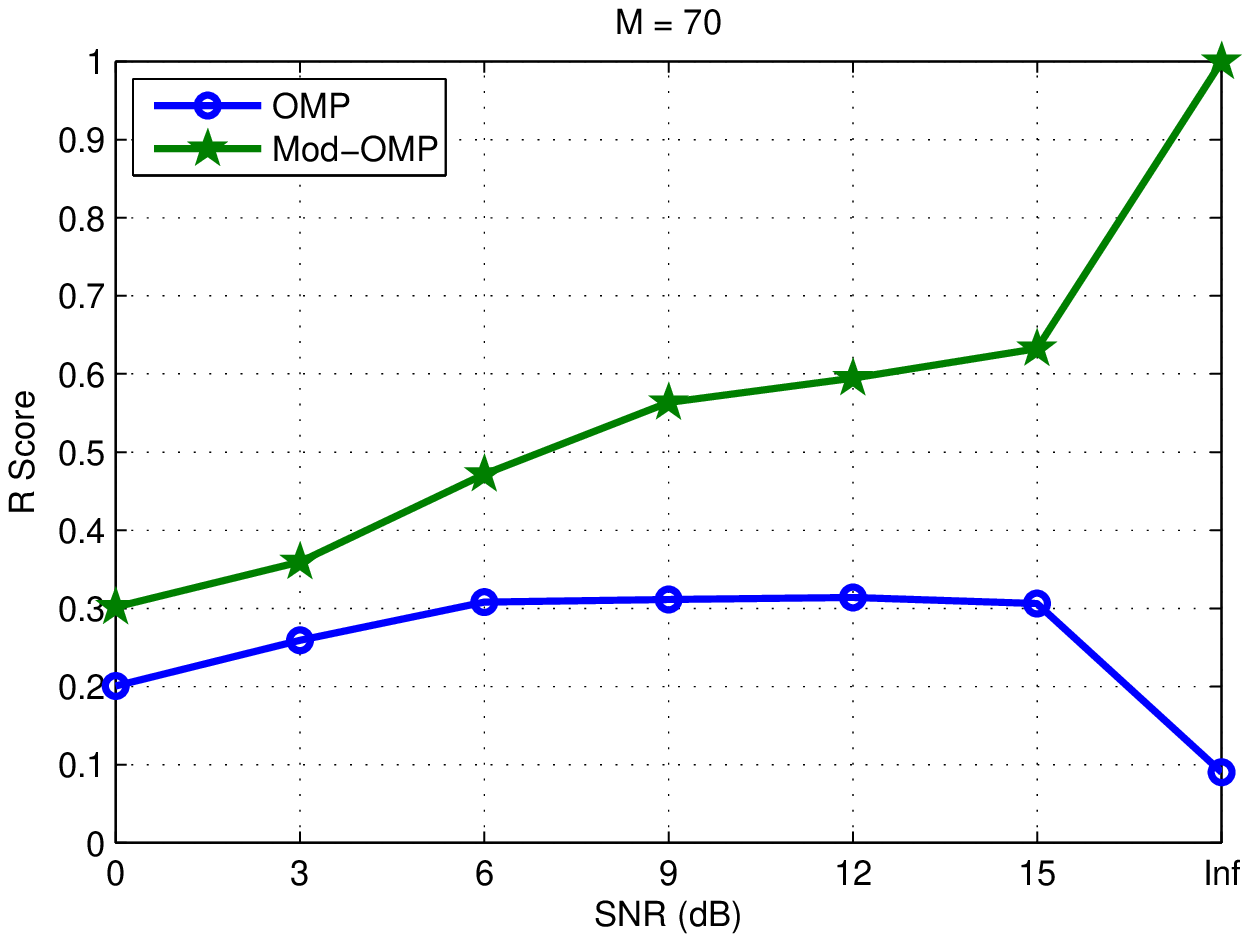}
	\caption{R Score vs. SNR for OMP and Mod-OMP in the single user WSS, $M = 70$.}
	\label{fig:M70_OMP_v_Mod}
	\end{minipage}
\end{figure}

Fig. \ref{fig:Dist_Spec_Sensing} depicts an example of a centralized CRN with 1 fusion center and 3 SUs. There are 3 PUs activated within the CRN. Each PU can only occupy the frequency band allocated to itself. The SUs in the CRN can use any of the unoccupied frequency bands. In the CRN, all SUs and the fusion center perform the CWSS to determine the temporarily unoccupied frequency bands. The bottom-right of Fig. \ref{fig:Dist_Spec_Sensing} shows the spectrum observed at each SU when 3 PUs are activated. Feature 4 can be observed, that the occupancy of the frequency bands are common among all the SUs. 
\subsection{Compressed Spectrum Sensing}
One challenge for the WSS is the sampling rate. Recent progress in compressed sensing \cite{donoho2006cs,candes2006cs} provides a framework to overcome the sampling rate problem. Here we briefly review the basic CSS models.
\subsubsection{Single User Compressed Spectrum Sensing Model}
Assume there is an analog signal $x\left( t \right)$ that can be represented by a finite weighted sum of basis functions $\psi_i\left(t\right)$ with coefficients ${s}\left(i\right)$:
\begin{equation}
\label{cs_x}
x\left( t \right) = \sum\limits_{i = 1}^N {{s}\left(i\right) \psi _i \left( t \right)} 
\end{equation}
In the discrete time framework, Equation \ref{cs_x} can be represented as ${\bf x} = {\bf \Psi s}$, where $\bf \Psi$ is the $N \times N$ matrix with each column as the discrete time representation of basis function $\psi_i\left(t\right)$, $\bf x$ is the $N \times 1$ vector, the discrete time representation of $x\left(t\right)$, and $\bf s$ is an $N \times 1$ signal vector with $N$ entries ${ s}\left(i\right)$. After sampling $\bf x$ with an $M \times N$ matrix $\bf \Phi$, we have the $N \times 1$ measurement vector $\bf y$:
\begin{equation}
\label{cs_y}
{\bf y = \Phi x = \Phi \Psi s = \Theta s}
\end{equation}
For convenience, we call $\bf \Theta$ the sampling matrix. If $\bf s$ is sparse, and if $\bf \Theta$ meets the restricted isometry property (RIP) \cite{donoho2006cs,candes2006cs}, then $\bf s$ can be reconstructed perfectly with $M < N$ measurements by solving the following $l_1$-norm optimization problem:
\begin{equation}
\label{cs_problem}
{\bf \hat s} = \arg \mathop {\min }\limits_{\bf s} \left\| {\bf s} \right\|_1  \ \ \ 
 s.t.\ \ {\bf y = \Theta s } \\ 
\end{equation}
where 
$\left\|  \bullet  \right\|_p$ is the $l_p$ norm defined as:
\begin{equation}
\label{p_norm}
\left\|  \bullet  \right\|_p  = \left( {\sum {\left(  \bullet  \right)^p } } \right)^{\frac{1}{p}} 
\end{equation}

In the CSS, $\bf s$ can be viewed as the spectrum observed at the SU, $\bf \Theta$ denotes the sampling method and $\bf y$ is the measurements at the receiver. If $\bf \Theta$ meets the RIP, $\bf s$ can be reconstructed perfectly via Equation \ref{cs_problem} with $M << N$ measurements, and therefore sub-Nyquist rate sampling can be achieved. The details for the design of $\bf \Theta$ can be found in \cite{AIC_2008,xampling_2009}. In this correspondence, however, we focus on the reconstruction algorithms solving Equation \ref{cs_problem}. We will integrate the specific prior knowledge from feature 2 -- feature 4 into OMP and introduce the Mod-OMP for decision fusion. 
\subsubsection{Cooperative Compressed Spectrum Sensing Model}
When $J$ signal vectors ${\bf s}_j$ are sampled with the same $\bf \Theta$ at each SU, we have the $M \times J$ measurement matrix ${\bf Y}$ and the $N \times J$ signal matrix $\bf S$:
\begin{equation}
\label{dcs_Y}
{\bf Y} = \left[ {\begin{array}{*{20}c}
   {{\bf y}_1 } & {{\bf y}_2 } &  \ldots  & {{\bf y}_J }  \\
\end{array}} \right] = {\bf \Theta }\left[ {\begin{array}{*{20}c}
   {{\bf s}_1 } & {{\bf s}_2 } &  \ldots  & {{\bf s}_J }  \\
\end{array}} \right] = {\bf \Theta S}
\end{equation}
Such sampling model is referred to the DCS \cite{wang2009distributed,DCS2005}. Similar to the single user CSS, $\bf S$ can be reconstructed from $\bf Y$ with $M << N$ measurements. The joint-sparsity-model 2 (JSM-2) is of special interest, where the index set $\left\{ {i|{\bf s}_j \left( i \right) \ne 0} \right\}$ are common for all ${\bf s}_j$ while with different values. Such commonness exactly fits feature 4, and SOMP has been adapted to reconstruct $\bf S$ \cite{wang2009distributed}. Again, we will focus on the reconstruction algorithm and introduce the novel Mod-SOMP for data fusion considering the specific prior knowledge provided from feature 2 -- feature 4.
\section{Cooperative Wideband Spectrum Sensing Algorithms}
\label{CWSS_Algorithms}
In this section, we describe the representations of the spectrum using feature 2 and summarize feature 3 in Table \ref{tab:category}. Then, the R score metric is reviewed. Finally, the Mod-OMP DeF and the Mod-SOMP DaF using feature 2 -- feature 4 will be introduced respectively. 
\subsection{Representation of the Frequency Allocation Information}
Assume a centralized CRN has $J$ SUs and a fusion center. The $B$ Hz spectrum available for the CRN is allocated with $K$ PUs. The discrete spectrum measured at the $j^{th}$ SU can be described using ${\bf s}_j$:
\begin{equation}
\label{SU_vector}
{\bf s}_j  = \left[ {\begin{array}{*{20}c}
   {s_j \left( {1 } \right)} & {s_j \left( {2} \right)} &  \cdots  & {s_j \left( {N } \right)}  \\
\end{array}} \right]^T 
\end{equation}
where $T$ is matrix transpose and $s_j\left( i\right)$ are the uniformly sampled frequency values over the $B$ Hz spectrum. According to feature 1, ${\bf s}_j$ is sparse:
\begin{equation}
\label{sparsity}
\left\| {{\bf s}_j } \right\|_0  /N \leq 10\%, \ \ j = 1,2,\cdots,J
\end{equation}

The frequency allocation information in feature 2 can be represented by index sets ${\bf u}_k$:
\begin{equation}
\label{band_alloc}
\begin{array}{l}
 {\bf u}_1  = \left\{ {1 ,2 , \cdots {b_1 } } \right\} \\ 
 {\bf u}_2  = \left\{ {{b_1  + 1} ,{b_1  + 2} , \cdots {b_2 } } \right\} \\ 
  \cdots  \\ 
 {\bf u}_K  = \left\{ {{b_{k - 1}  + 1} ,{b_{k - 1}  + 2} , \cdots N } \right\} \\ 
 \end{array}
\end{equation}
where ${b_1} \cdots {b_{k-1}}$ are frequency indices determining the boundaries of different frequency bands.

Feature 3 is summarized in Table \ref{tab:category} via time and frequency. ${\bf u}_k$ can be grouped into the category sets ${\bf C}_n$:
\begin{equation}
\label{subsets}
\begin{array}{l}
 {\bf C}_n  = \left\{ {{\bf u}_k \left| {k \in {\rm Category }\;n} \right.} \right\},n = 1,2,3 \\ 
 {\bf \Omega } = \left\{ 1,2,\cdots,N \right\} = \bigcup\limits_n {{\bf C}_n } \; \; \left( {\bf C}_i  \cap {\bf C}_j  = \emptyset ,{\rm  for }\;i \ne j \right)\\ 
 \end{array}
\end{equation}
In the noiseless case, if $\left\| {s_j \left( {i } \right)} \right\| > 0$ and if $i  \in {\bf u}_k$, then frequency band ${\bf u}_k$ is occupied and all $i$ within ${\bf u}_k$ should not be used by the CRN. Let the $K \times 1$ vector $\bf o$ denote the occupancy decisions of the frequency bands. If ${\bf u}_k$ is occupied, then ${\bf o}\left(k\right) = 1$, otherwise ${\bf o}\left(k\right) = 0$. 

If decision fusion is used, each SU will first obtain a local estimate ${\bf \hat o}_j$, then the fusion center will obtain an overall estimate ${\bf \hat o}$. For data fusion, $\bf \hat o$ is obtained at the fusion center directly. 

Since there are $K$ decisions to be made for the whole spectrum, we need to consider the following decision statistics \cite{chen2004statistical,Meng2011CWSS} from $\bf o$ and $\bf \hat o$:
\begin{itemize}
	\item Number of Hit, $a$, the number of correct decisions for ${\bf o}(k) = {\bf \hat o}(k) = 1$.
	\item Number of Miss, $b$, the number of wrong decisions for ${\bf o}(k) = 1, \ \ {\bf \hat o}(k) = 0$.
	\item Number of False, $c$, the number of wrong decisions for ${\bf o}(k) = 0, \ \ {\bf \hat o}(k) = 1$.
	\item Number of Correct, $d$, the number of correct decisions for ${\bf o}(k) = {\bf \hat o}(k) = 0$.
\end{itemize}

The Hanssen-Kuiper or R score is a common metric using the above statistics to evaluate the performance of correct decisions and wrong decisions. The R score is calculated as:
\begin{equation}
\label{R_score}
R = w_1 \frac{a}{{a + b}} - w_2 \frac{c}{{c + d}}
\end{equation}
where $w_1$ and $w_2$ are the weights denoting the credit to correct decisions and the penalty to wrong decisions, respectively. In this correspondence, we set $w_1 = w_2 = 1$, meaning that the loss due to wrong decisions is equal to the benefit from correct decisions. Other weights can be set according to desired situations. In such setup, $R \in \left[ { - 1,1} \right]$, with $1$ as the score indicating that all decisions are correct, while $-1$ as the score indicating that all decisions are wrong.
Next, we will modify OMP and SOMP with the prior knowledge ${\bf u}_k$ and ${\bf C}_n$. 
\subsection{The Mod-OMP for Decision Fusion}
Mod-OMP is modified upon the original OMP. For the $j^{th}$ SU, Mod-OMP starts with local measurement vector ${\bf y}_j$, sampling matrix $\bf \Theta$, error tolerance $\eta$ determined by noise level, and the index sets ${\bf u}_k$ and ${\bf C}_k$. Mod-OMP gets the estimate ${\bf \hat s}_j$ from the following steps:
\begin{enumerate}
	\item \textbf{Initialize:} Set the iteration counter $t = 0$. Initialize the residual ${\bf res}^{(t)}_j= {\bf y}_j$, ${\bf \Theta^{(t)}} = \bf \Theta$, the index set ${\bf \Lambda}^{(t)} = {\bf C}_1$. If ${\bf \Lambda}^{(t)} = \emptyset$, jump to Step 4.
	\item Solve the least-squares problem to obtain ${\bf z}_j$:
\begin{equation}
\label{leastsquares}
{\bf z}_j  = \arg \mathop {\min }\limits_{\bf z} \left\| {{\bf \Theta }^{(t)}{\bf z} - {\bf y}_j} \right\|_2 
\end{equation}
	\item Update the new residual:
\begin{equation}
\label{getresidual}
{\bf res}^{(t)}_j = {\bf y}_j - {\bf \Theta }^{(t)} {\bf z}_j
\end{equation}
	\item Increment $t$. Find the index $\lambda^{(t)}$ that satisfies the following equation:
\begin{equation}
\label{lambdamax}
\lambda ^{(t)}= \arg \mathop {\max }\limits_{i = 1,...,N} \left| {\left\langle {{\bf res}^{(t-1)}_j ,{\bf \theta }_i } \right\rangle } \right|
\end{equation}
where ${\bf \theta}_i$ denotes the $i^{th}$ column vector of ${\bf \Theta}$. $\left\langle {{\bf \alpha,\beta}} \right\rangle$ denotes the inner product of two vectors $\bf \alpha$ and $\bf \beta$. If the maximum occurs for multiple indices, break the tie deterministically.
	\item If $\lambda^{(t)} \in {\bf u}_k \subset {\bf C}_2$, let ${\bf \Lambda}^{(t)}= {\bf \Lambda}^{(t-1)}\bigcup {\bf u}_k$. If $\lambda^{(t)}\in {\bf u}_k \subset {\bf C}_3$ or ${\bf C}_n = \emptyset$, let ${\bf \Lambda}^{(t)} = {\bf \Lambda}^{(t-1)}\bigcup \lambda^{(t)}$. 
	\item Set the matrix of chosen atoms ${\bf \Theta}^{(t)}= {\bf \Theta} _{{\bf \Lambda}^{(t)}}$, where ${\bf \Theta} _{{\bf \Lambda}^{(t)}}$ is a sub-matrix of ${\bf \Theta}$ containing columns with indices in ${{\bf \Lambda}^{(t)}}$.
	\item Obtain ${\bf z}_j$ using Equation \ref{leastsquares} and update the new residual ${\bf res}^{(t)}_j$ using Equation \ref{getresidual}.
	\item If $t < m$ or $\left\| {{\bf res}^{(t)}_j} \right\|_2  > \eta $, return to Step 4. Otherwise, ${\bf z}_j$ is the estimated signal ${\bf \hat s}_j$ with non-zeros at indices listed in ${\bf \Lambda}^{(t)}$. 
\end{enumerate}

Notice that if ${\bf C}_n = \emptyset$ and ${\bf u}_k = \emptyset$, there is no prior knowledge available, and Mod-OMP becomes the original OMP. It can be seen that OMP only finds one non-zero index $i$ in each iteration and treats all $i$ independently. The modifications made by Mod-OMP using the prior knowledge ${\bf u}_k$ and ${\bf C}_n$ are described as follows. $i \in {\bf C}_1$ are included in the initial index set as ${\bf \Lambda}^{(0)}= {\bf C}_1$, because values under those indices are non-zero all the time. For $i \in {\bf C}_2$, if a single $i$ is selected in iteration $t$ and if $i \in {\bf u}_k$, then ${\bf \Lambda}^{(t)}= {\bf \Lambda}^{(t-1)} \bigcup {\bf u}_k$, because there must be other non-zeros values under frequency band ${\bf u}_k$. For $i \in {\bf C}_3$, only one $i$ will be selected in each iteration, because frequency bands in ${\bf C}_3$ are only partially occupied.

The modifications above have the following advantages:
\begin{enumerate}
	\item The number of iterations are reduced from the number of occupied frequency points to the number of occupied frequency bands.
	\item The reconstruction errors are restricted within several selected frequency bands. For the original OMP, however, the reconstruction errors are spread out in any frequency bands. 
\end{enumerate}

When ${\bf \hat s}_j$ is reconstructed at the $j^{th}$ SU by OMP/Mod-OMP, a local occupancy estimate ${\bf \hat o}_j$ needs to be obtained. For the $k^{th}$ frequency band, ${\bf \hat o}_j\left(k\right) = 1$ if at least one non-zero spike is detected. Otherwise ${\bf \hat o}_j\left(k\right) = 0$:
\begin{equation}
\label{SU_decision_rule}
\max \left\| {{\bf \hat s}_j \left( i \right)} \right\| > T_f \ \ and \ \ i \in u_k
\end{equation}
where $T_f$ is the threshold determined by the noise level. 

The fusion center then obtains ${\bf \hat o}$ from all ${\bf \hat o}_j$. According to feature 4, the occupied frequency bands are common among all SUs. Though each SU estimates ${\bf \hat o}_j$ independently, the original ${\bf o}_j$ are the same. Therefore, for frequency band $k$, we can use the $L$ out of $J$ rule for decision making \cite{zeng2010review}:
\begin{equation}
\label{FC_decision_rule}
\sum\limits_{j = 1}^J {{\bf \hat o}_j \left( k \right)}
\substack{{\mathop  > \limits^{{\bf \hat o}\left( k \right) = 1} } \\ {\mathop  < \limits_{{\bf \hat o}\left( k \right) = 0} }} L
\end{equation}
In this correspondence, we set $L = \left\lceil {\frac{J}{2}} \right\rceil $, where $\left\lceil  \bullet  \right\rceil$ is the smallest integer not less than $\bullet$. This means that ${\bf o}\left(k\right) = 1$ \textsl{iff} more than half of the SUs make the decision that frequency band $k$ is occupied. Such rule can effectively reduce the number of wrong decisions, because each SU makes decisions independently, and the probability for $L$ SUs to make the same wrong decisions is low.

In summary, the OMP/Mod-OMP DeF gets the estimate $\bf \hat o$ by the following steps:
\begin{enumerate}
	\item Each SU reconstructs the spectrum ${\bf \hat s}_j$ from ${\bf y}_j$ by OMP/Mod-OMP.
	\item Each SU uses ${\bf \hat s}_j$ to estimate an occupancy vector ${\bf \hat o}_j$ and sends it to the fusion center.
	\item The fusion center uses ${\bf \hat o}_j$ to determine ${\bf \hat o}$ according to the $L$ out of $J$ rule.
\end{enumerate}	
\subsection{The Mod-SOMP for Data Fusion}
The proposed Mod-SOMP is modified upon the original SOMP. Mod-SOMP starts with $\bf Y$, $\bf \Theta$, $\eta$, ${\bf u}_k$ and ${\bf C}_n$. Mod-SOMP gets the estimate ${\bf \hat S}$ as the following:
\begin{enumerate}
	\item \textbf{Initialize:} Set the iteration counter $t = 0$. Initialize the residual ${\bf Res}^{(t)} = {\bf Y}$, ${\bf \Theta^{(t)}} = \bf \Theta$, the index set ${\bf \Lambda}^{(t)} = {\bf C}_1$. If ${\bf \Lambda}^{(t)} = \emptyset$, jump to Step 4.
	\item Obtain ${\bf Z} = \left[ {\bf z}_1, {\bf z}_2, \cdots {\bf z}_J \right]$ using Equation \ref{leastsquares}.
	\item Update the new residual matrix ${\bf Res}^{(t)}  = \left[ {\begin{array}{*{20}c}
   {{\bf res}^{(t)}_1 } & {{\bf res}^{(t)}_2 } &  \ldots  & {{\bf res}^{(t)}_J }  \\
\end{array}} \right] $ using Equation \ref{getresidual}.
	\item Increment $t$. Find the index $\lambda^{(t)}$ that satisfies the following equation:
\begin{equation}
\label{sum_lambdamax}
\lambda ^{(t)}= \arg \mathop {\max }\limits_{i = 1,2, \ldots N} \sum\limits_{j = 1}^J {\left| {\left\langle {{{\bf res}^{(t)}_j} ,\theta _i } \right\rangle } \right|} 
\end{equation}
	\item If $\lambda^{(t)} \in {\bf u}_k \subset {\bf C}_2$, let ${\bf \Lambda}^{(t)}= {\bf \Lambda}^{(t-1)}\bigcup {\bf u}_k$. If $\lambda^{(t)}\in {\bf u}_k \subset {\bf C}_3$ or ${\bf C}_n = \emptyset$, let ${\bf \Lambda}^{(t)} = {\bf \Lambda}^{(t-1)}\bigcup \lambda^{(t)}$. 
	\item Set the matrix of chosen atoms ${\bf \Theta}^{(t)}= {\bf \Theta} _{{\bf \Lambda}^{(t)}}$. 
	\item Obtain ${\bf Z}$ using Equation \ref{leastsquares} and update the new residual ${\bf Res}^{(t)}$ using Equation \ref{getresidual}.
	\item If $t < m$ or $\left\| {\bf Res}^{(t)}\right\|_2 > \eta$, return to Step 4. Otherwise, $\bf Z$ is the estimated signal matrix $\bf \hat S$ with non-zero rows listed in ${\bf \Lambda}^{(t)}$.
\end{enumerate}

Mod-SOMP becomes the original SOMP when ${\bf C}_n = \emptyset$ and ${\bf u}_k = \emptyset$. SOMP finds one non-zero index $i$ in each iteration and treats all $i$ independently. Mod-SOMP treats $i$ with prior knowledge in the same way as Mod-OMP. Like Mod-OMP, Mod-SOMP has the same advantages over the original SOMP.

When $\bf \hat S$ is obtained, the fusion center first sums the columns of $\bf \hat S$ to get $\bf \hat s$ \cite{zeng2010review}:
\begin{equation}
\label{overall_s}
{\bf \hat s}\left( i \right) = \sum\limits_{j = 1}^J {\left\| {{\bf \hat S}\left( {i,j} \right)} \right\|} 
\end{equation}
For the $k^{th}$ frequency band, ${\bf \hat o}\left(k\right) = 1$ if the following statement is true. Otherwise ${\bf \hat o}\left(k\right) = 0$.
\begin{equation}
\label{FC_decision_rule}
\max \left\| {{\bf \hat s} \left( i \right)} \right\| > T_f \ \ and \ \ i \in u_k
\end{equation}

In summary, the SOMP/Mod-SOMP DaF gets the estimate $\bf \hat o$ by the following steps:
\begin{enumerate}
	\item Each SU collects its own measurement vector ${\bf y}_j$ and sends it to the fusion center, forming $\bf Y$.
	\item The fusion center uses $\bf Y$ to reconstruct an estimate of the signal matrix $\bf \hat S$ by SOMP/Mod-SOMP.
	\item The fusion center uses ${\bf \hat S}$ to determine ${\bf \hat o}$.
\end{enumerate}
\section{Simulation Results}
\label{Simulation_Results}
In this section, we will use the Monte Carlo simulation to compare the performances of the OMP DaF, Mod-OMP DaF, SOMP DeF, and the Mod-SOMP DeF. $500$ simulations are conducted for each setup. In the simulations, there are $K = 41$ frequency bands allocated to the PUs over a discrete spectrum with $N = 200$. The frequency allocation information is defined by ${\bf u}_k$ and ${\bf C}_n$. There are $2$ ${\bf u}_k$ in ${\bf C}_1$, and $\left\| {{\bf C}_1 } \right\|_0 = 5$; $33$ ${\bf u}_k$ in ${\bf C}_2$ and $\left\| {{\bf C}_2 } \right\|_0 = 165$; $6$ ${\bf u}_k$ in ${\bf C}_3$ and $\left\| {{\bf C}_3 } \right\|_0 = 30$. These numbers can be set according to the specific frequency allocation information in a specific spectrum range.

In the simulation, a total of $10$ ${\bf u}_k$ are occupied randomly, with $2$ in ${\bf C}_1$, $2$ in ${\bf C}_2$ and $6$ in ${\bf C}_3$. The occupancy of the whole spectrum is around $10\%$ ($20$ non-zero indices). Independent Rayleigh fading effects are applied at each SU, as well as Gaussian noise.

We first simulate the $J = 1$, the single user case, with $M = 70$. 
In this setup, only OMP and Mod-OMP are compared. The R score vs. SNR plots are depicted in Fig. \ref{fig:M70_OMP_v_Mod}. From the R score we can see that the $M = 70$ is insufficient for OMP to get perfect correct decisions even when SNR is infinity. The R score for Mod-OMP, however, is increasing with SNR, and reaches 1 when SNR is infinity.

\begin{figure}[ht]
	\begin{minipage}[b]{0.5\linewidth}
		\centering
		\includegraphics[width=1.0\textwidth]{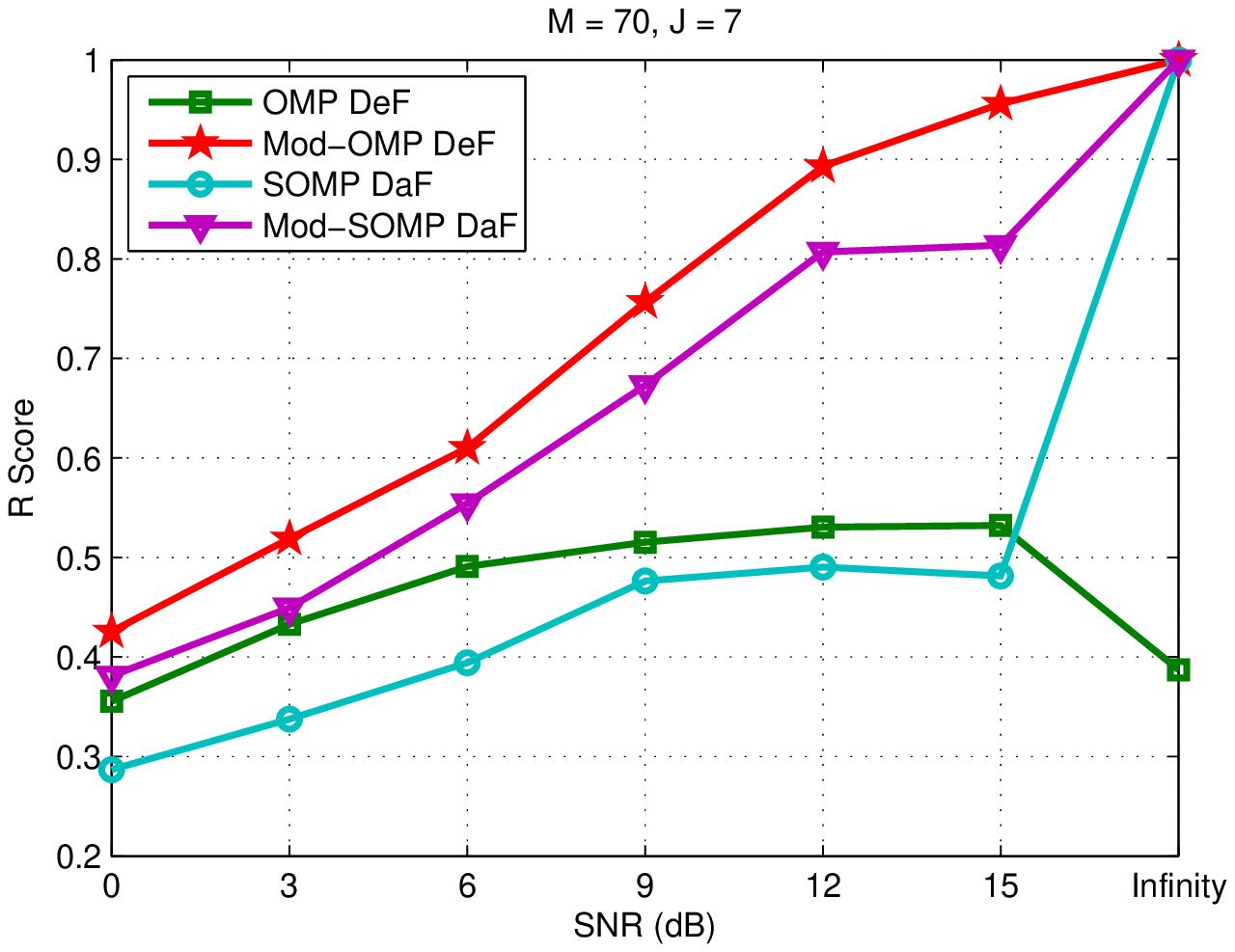}
		\caption{R Score vs. SNR for the OMP DeF, Mod-OMP DeF, SOMP DaF and the Mod-SOMP DaF in the CWSS. $M = 70, J = 7$.}
		\label{fig:M70_J7_All}
	\end{minipage}
\hspace{0.5cm}
	\begin{minipage}[b]{0.5\linewidth}
	\centering
		\includegraphics[width=1.0\textwidth]{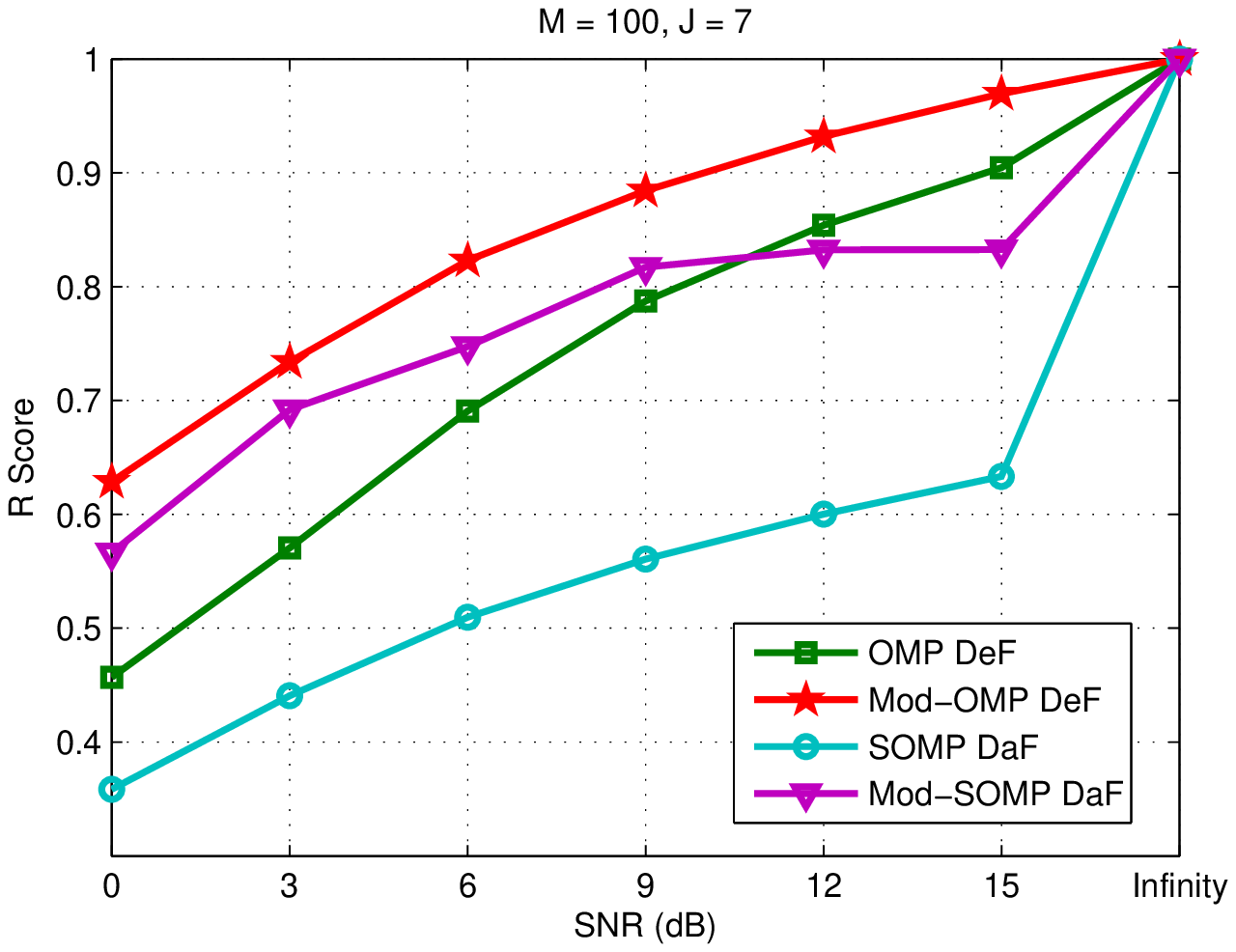}
		\caption{R Score vs. SNR for the OMP DeF, Mod-OMP DeF, SOMP DaF and the Mod-SOMP DaF in the CWSS. $M = 100, J = 7$.}
		\label{fig:M100_J7_All}
	\end{minipage}
\end{figure}

Then, we simulate with $J = 7$ for all 4 CWSS algorithms with $M = 70$. The R score vs. SNR plots are depicted in Fig. \ref{fig:M70_J7_All} and Fig. \ref{fig:M100_J7_All}, with $M = 70$ and $M = 100$, respectively. From Fig. \ref{fig:M70_J7_All}, it can be seen that, the OMP DeF still cannot give perfect correct decisions when SNR is infinity. This is because none of the SU is able to provide correct decisions. The R scores of the other three algorithms, however, are able to reach $1$ when SNR is infinity. The Mod-OMP DeF is better than the Mod-SOMP DaF, while the Mod-SOMP DaF is better than the SOMP DaF. This shows the advantage of using prior knowledge.

From Fig. \ref{fig:M100_J7_All}. We can see that with increased number of samples $M$, all algorithms have relatively better performance. The R scores for all 4 algorithms can reach $1$ when SNR is infinity. The Mod-OMP DeF always has the highest R score. The Mod-SOMP DaF has higher R score than the OMP DeF when SNR is low, while lower R score than the OMP DeF when SNR is high. The SOMP DaF has lowest R score for all SNR. Generally speaking, the decision fusion algorithms (the OMP/Mod-OMP DeF) have higher R score than their data fusion counterparts (the SOMP/Mod-SOMP DaF). This is because decision fusion makes independent decisions in each SU and has the $L$ out of $J$ rule to suppress wrong decisions.
	
Notice that the SOMP DaF has very poor performance. We analyze one example of the decision statistics in Table \ref{Decision_Statistics} for $M = 100, J = 7$ and SNR = $15$ dB. If all decisions are correct, $a = 10$, $b = 0$, $c = 0$ and $d = 31$. It can be seen that thanks to the $L$ out of $J$ rule, the decision fusion based algorithms make fewer wrong decisions on both occupied frequency bands and unoccupied frequency bands. The data fusion algorithms, however, cannot suppress wrong decisions on the unoccupied frequency bands. 
Then, it can be seen that both Mod-OMP DeF and Mod-SOMP DaF with the prior knowledge have less wrong decisions on the unoccupied frequency bands than their counterparts without the prior knowledge. This is because when the prior knowledge is applied, the algorithm will select one frequency band in each iteration, and the errors are restricted within the selected bands only. The algorithms without the prior knowledge, however, do not have such property and errors can be spread out in all frequency bands. 
\subsection{Discussion}
A centralized CRN can choose between decision fusion and data fusion, depending on the required computation burden and communication burden. Decision fusion has high computation burden on each SU but low computation on the fusion center. Meanwhile, it has few communication burden because each SU only sends a binary vector ${\bf o}_j$. Data fusion, however, has almost no computation burden on each SU but high computation burden on the fusion center. The communication burden is higher than that of decision fusion, because each SU needs to send the quantized sample vector ${\bf y}_j$. 
\section{Conclusions}
\label{Conclusions}
In this correspondence, we have summarized four specific features of the CWSS and shown how to integrate those features in the algorithm design. With feature 1, we are able to use compressed sensing. With feature 2 and feature 3 as the prior knowledge, we are able to modify OMP and SOMP and proposed the Mod-OMP and Mod-SOMP. Then, with feature 4, we are able to integrate those algorithms in a centralized CRN. Four CWSS algorithms have been introduced, the OMP DeF, Mod-OMP DeF, SOMP DaF the Mod-SOMP DeF. Their decision performances in the CWSS have been compared by Monte Carlo simulations using the R score metric. From the simulation results, we have verified that the minimum number of measurements to achieve error free decisions can be reduced when the prior knowledge is applied. Then, we have verified that the decision fusion based OMP DeF and Mod-OMP DeF have outstanding wrong decision suppression on both occupied frequency bands and unoccupied frequency bands. Finally, the Mod-OMP DeF and the Mod-SOMP DaF with the prior knowledge can restrict wrong decisions on unoccupied frequency bands. 

\bibliographystyle{ieeetrans}
\bibliography{bib/CR,bib/CR_CS,bib/Qiu_Group_bib,bib/Compressed_sensing/Applications/Analog_Information_Conversion,bib/Compressed_sensing/Applications/Astronomy,bib/Compressed_sensing/Applications/Biosensing,bib/Compressed_sensing/Applications/Communications,bib/Compressed_sensing/Applications/Compressive_Imaging,bib/Compressed_sensing/Applications/Compressive_Radar,bib/Compressed_sensing/Applications/Geophysical_Data_Analysis,bib/Compressed_sensing/Applications/Hyperspectral_Imaging,bib/Compressed_sensing/Applications/Medical_Imaging,bib/Compressed_sensing/Applications/Spectrum_analysis,bib/Compressed_sensing/Applications/Surface_metrology,bib/Compressed_sensing/Compressive_Sensing/compressive_sensing,bib/Compressed_sensing/Recovery_Algorithms/recovery_algorithms,bib/Compressed_sensing/Data_Stream_Algorithms/Dimension_Reduction_Embeddings,bib/Compressed_sensing/Data_Stream_Algorithms/Heavy_Hitters,bib/Compressed_sensing/Data_Stream_Algorithms/Histogram_Maintenance,bib/Compressed_sensing/Data_Stream_Algorithms/Random_Sampling,bib/Compressed_sensing/Extensions/extensions,bib/Compressed_sensing/Extensions/multi_sensor_distributed_CS,bib/Compressed_sensing/Foundations_Connections/Bayesian_Methods,bib/Compressed_sensing/Foundations_Connections/Coding_Info_Thry,bib/Compressed_sensing/Foundations_Connections/Finite_Rate_Innovation,bib/Compressed_sensing/Foundations_Connections/High_Dim_Geometry,bib/Compressed_sensing/Foundations_Connections/l1_Norm_Minimization,bib/Compressed_sensing/Foundations_Connections/Lossy_Compression,bib/Compressed_sensing/Foundations_Connections/Machine_Learning,bib/Compressed_sensing/Foundations_Connections/Multiband_Signals,bib/Compressed_sensing/Foundations_Connections/Statistical_Signal_Processing,bib/Compressed_sensing/Tutorials/TUTORIALS}

\begin{table}[htp]
\begin{minipage}[b]{0.5\linewidth}
\centering
\caption{Properties of the Frequency Band Occupancy}
\label{tab:category}
\begin{tabular}{|r|r|r|}
\hline
	& Time & Frequency	\\
\hline
Category 1 & Always Occupied & Fully Occupied	\\
\hline
Category 2 & Sometimes Occupied & Fully Occupied	\\
\hline
Category 3 & Sometimes Occupied & Partially Occupied	\\
\hline
\end{tabular}
\end{minipage}
\hspace{0.5cm}
\begin{minipage}[b]{0.5\linewidth}
\centering
\caption{Decision Statistics for All Algorithms}
\label{Decision_Statistics}
\begin{tabular}{|r|r|r|r|r|}
\hline
 & Mod-OMP DaF & OMP DaF & Mod-SOMP DeF & SOMP DeF\\
\hline
$a$ & 9.966 & 9.982 & 10 & 10\\
\hline
$b$ & 0.034 & 0.018 & 0 & 0\\
\hline
$c$ & 0.842 & 2.906 & 5.194 & 11.944\\
\hline
$d$ & 30.158 & 28.094 & 25.806 & 19.056\\
\hline
\end{tabular}
\end{minipage}
\end{table}

%
%
%
%
%
%

%
\IEEEpeerreviewmaketitle

\end{document}